\documentclass[preprint,prl,aps]{revtex4}
\usepackage{amsmath,graphicx}

\begin{document}

\title{$e^+e^-$ Pair Production from 10 GeV to 10 ZeV}

\author{Spencer R. Klein}

\address{Lawrence Berkeley National Laboratory, Berkeley, CA 94720, USA}

\begin{abstract}

At very high energies, pair production ($\gamma\rightarrow e^+e^-$)
exhibits many interesting features.  The momentum transfer from the
target is very small, so the reaction probes the macroscopic
properties of the target, rather than individual nuclei.  Interference
between interactions with different atoms reduces the pair production
cross section considerably below the Bethe-Heitler values.  At very
high energies, photonuclear interactions may outnumber pair
production.

In contrast, in crystals, the interaction amplitudes may add
coherently, greatly increasing the cross sections.  Pair production in
matter-free magnetic fields is also possible.  The highest energy pair
production occurs at high energy particle colliders.  This article
will compare pair production in these very different regimes.

\end{abstract}

\maketitle
\narrowtext

\section{Introduction}

Pair production was first observed by Anderson in 1932, who used the
process to discover the positron \cite{anderson}.  Our modern
theoretical understanding of pair production stems from work by Bethe
and Heitler \cite{bh}.  At high photon energies, the pair production
cross section reaches a constant value \cite{greisen},
\begin{equation}
\sigma = {7A\over 9X_0N_A},
\label{eq:constsigma}
\end{equation}
where $X_0$ is the target radiation length, in units of column density
(mass/length$^2$, usually g/cm$^2$), $A$ is the atomic mass of the
target (usually in g/mole), and $N_A = 6.022\times10^{23}$ is
Avogadro's number \cite{pdg}.  $X_0$ is also commonly given in units
of length; this length can be multiplied by the density $\rho$ to give
$X_0$ in column density.

Eq. \ref{eq:constsigma} applies at high photon energies, $k > 10$ GeV,
when the momentum transfer from the target to the pair is sufficiently
small.  The atomic electrons screen the nucleus from the incident
photon/pair, limiting the effective impact parameter to the
Thomas-Fermi radius.  In this complete screening limit, the cross
section is independent of photon energy.  The differential cross
section is usually given in terms of $x=E/k$, where $E$ is the
electron energy \cite{skreview}:
\begin{equation}
{d\sigma\over dx} = {A\over X_o N_A} \bigg[1-{4\over 3}x(1-x)\bigg].
\label{eq:dsde}
\end{equation}
This lowest order cross section is symmetric in $x$ and $1-x$, the
electron and positron energy.  

At high energies, two regimes are of interest.  Most conversions
produce pairs near threshold, with pair mass $M_p\approx 2 m$, where
$m$ is the electron mass.  The momentum transfer required to produce a
pair decreases linearly as the photon energy rises.  The longitudinal
momentum transfer required to produce a lepton pair with energies $E$
and $k-E$ is
\begin{equation}
q_{||} = k - \sqrt{(k-E)^2-m^2} - \sqrt{E^2-m^2} \approx {m^2 k\over 2E(E-k)}
\approx {M_p^2\over 2k}
\label{eq:qperp0}
\end{equation}
We take $c=1$ throughout this review.  The approximation in the last
two formulae cover are accurate for the vast majority of the cross
section, but fail for very asymmetric pairs, where $x\rightarrow 0$ or
$x\rightarrow 1$.

As $k$ rises and $q_{||}$ drops, the uncertainty principle requires
that the location of the conversion becomes more and more delocalized.
The pair production is delocalized over a region known as the
formation length:
\begin{equation}
l_{f0} ={\hbar\over q_{||}} = {2\hbar E(k-E)\over m^2k} = {2\hbar k\over
M_p^2}.
\label{eq:lf0}
\end{equation}
The subscript '0' shows that this is for pair conversion in free
space; future sections will consider modifications to $l_f$ in dense
media.  The formation length grows linearly with the photon energy.
When the formation length is larger than the typical inter-nuclear
separation, interactions with different atoms are no longer
independent.

This loss of independence leads to many interesting consequences.  In
amorphous materials, the pair production cross section decreases as
the photon energy rises, invalidating Eqs. \ref{eq:constsigma} and
\ref{eq:dsde}.  The conversion probability is no longer proportional
to the number of target atoms.  In moderately thick targets, the
conversion probability may even be proportional to the logarithm of
the thickness!  The reduced cross sections appear in a number of 'real
world' situations.  For example, they affect electromagnetic showers
produced by very high energy cosmic rays.

In crystals, the long formation length allows pair conversion
amplitudes from regularly spaced atoms to add coherently. The strong
magnetic fields can greatly increase the pair production cross
section.  This enhancement may be useful in accelerator design.

\begin{figure}[bt]
\center{\includegraphics[angle=270,clip=,width=0.8\columnwidth]{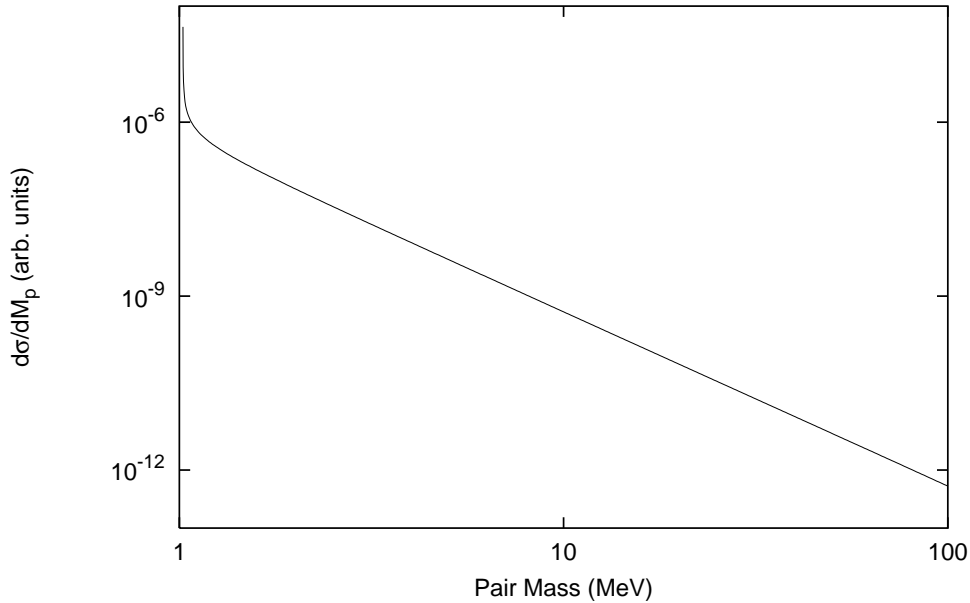}}
\caption[]{$d\sigma/dM_p$ for pair production from a 100 GeV photon.
The cross section is strongly peaked near threshold.  Away from the
peak, the cross section scales as $1/M_p^3$.  The spectrum is almost
independent of the photon energy.  For high enough pair masses, the
interaction is sensitive to the targets internal structure, and the
cross section drops falls off more rapidly.}
\label{fig:bhwide}
\end{figure}

For very-high-mass pairs, the momentum transfer from the pair is high,
and production of these pairs can probe short distance scales.
High-mass pairs have been used to test quantum electrodynamics (QED).
Neglecting the transverse momentum of the pair (a relatively small
effect), the pair mass is
\begin{equation}
M_p^2 = {m^2 \over x(1-x)}.
\label{eq:xx}
\end{equation}
Figure \ref{fig:bhwide} shows the pair mass spectrum, $d\sigma/dM_p$.
The cross section is strongly peaked near threshold, $x\approx 0.5$
and $M_p\approx 2m$, In this regime, the electron and positron have a
small (non-relativistic) relative velocity, and the cross section is
enhanced.  At even slightly higher masses, the differential velocity
is large and the cross section scales as
\begin{equation}
{d\sigma\over dM_p} \propto {1\over M_p^3}.
\label{eq:dsdm}
\end{equation}
This $1/M_p^3$ dependence continues as long as the momentum transfer
to the target is small.  When the momentum transfer rises, screening
is no longer complete, and the structure of the target becomes
significant.  In this regime, the cross section decreases faster than
$1/M_p^3$.

More recently, QED has been studied at particle colliders, which can
reach shorter distance scales.  Pair production reactions such as
$e^+e^-\rightarrow e^+e^-e^+e^-$ are an important channel for these
studies.  Another important channel for testing QED has been pair
production at heavy ion colliders.  Here, the very strong fields are
of interest.  For gold or lead, the perturbative expansion coupling
$Z\alpha\approx 0.6$, and it is not unreasonable to expect significant
non-perturbative effects in heavy ion collisions.  Here, $\alpha =
e^2/\hbar c\approx 1/137$ is the electromagnetic coupling constant,
with $e$ the charge on the electron.

High energy electrodynamics in matter has been previously reviewed by
Akhiezer and Shulga \cite{aksreview}, Baier and Katkov
\cite{bkreview}, and by Klein \cite{skreview}.  Although it does not
address pair production, Ter Mikaelians book \cite{termikaelian} has
a very interesting, largely classical discussion of high energy
bremsstrahlung. Interactions in crystals have been considered by
Palazzi \cite{palazzi}, S\o rensen \cite{sorensen}, and by Baier,
Katkov and Strakhovenko \cite{bkscrystal}.  Pair production in heavy
ion collisions has been recently reviewed by Gerhard Baur and his
collaborators \cite{baurreview}.

Section 2 of this review considers pair production from isolated
atoms.  Section 3 will discuss pair production in bulk media.  Section
4 will cover production in discrete fields and crystals.  Section 5
will consider pair production at colliding beam accelerators, and
Sec. 6 will draw some conclusions.

\section{Pair Production from Isolated Atoms}

\begin{figure}[bt]
\center{\includegraphics[clip=,width=0.9\columnwidth]{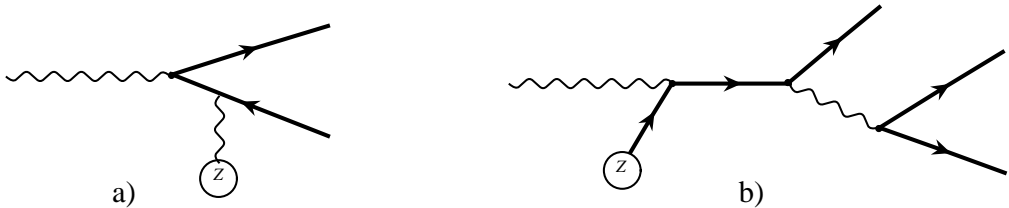}}
\caption[]{A schematic diagram of (a) direct pair production on a
target with charge $Z$ and (b) pair production via the Compton
process.  Diagram (a) is usually dominant, and the Compton process (b)
is mainly visible via it's interference with direct production.}
\label{fig:scheme}
\end{figure}

Pair production from an isolated atom occurs schematically via a
two-photon process, as is shown in Fig. \ref{fig:scheme}(a); In the
target frame of reference, the incident photon fluctuates to a virtual
$e^+e^-$ pair.  The pair combines with a virtual photon from the
target to form a real $e^+e^-$ pair. In the competing Compton process,
Fig. \ref{fig:scheme}(b), the photon is first absorbed by a nucleus or
electron, which then emits a virtual photon, which decays to a
$e^+e^-$ pair.  The Compton amplitude is relatively small, and the
process is usually observed via it's interference with the direct
reaction \cite{bh}.  This review will focus on the kinematics of pair
production, which determines many characteristics of the reaction.

As long as $l_f > a$, where the Thomas-Fermi atomic radius $a= 0.8
Z^{-1/3}a_0$ ($a_0= 5.3$ nm is the Bohr radius) then the nascent pair
couples to the target atom as a whole.  This is the complete screening
limit; at distances larger than $a$, the electrons screen the nuclear
charge.  In this limit, the cross section for pair production is
independent of $k$.  For heavy ions (heavier than iron), the limit is
reached within 1\% for $k>10$ GeV; for lighter ions, higher photon
energies are required.  For hydrogen, the finite-energy correction is
still slightly larger than 1\% for $k=100$ GeV \cite{tsai}.

For light ions, the radiation length $X_0$ scales as 1/$Z^2$ - the
interaction amplitude depends on $Z$, and the cross section on $Z^2$.
However, as $Z$ rises, higher order diagrams with multiple photons may
be important.  The most important (largest) higher order terms account
for the fact that the pair is produced in the potential well of the
target nucleus; the other terms, due to fluctuations in the incoming
photon, are small.  Bethe and Maximon calculated the corrections due
to the potential by solving exactly the Dirac equation for an electron
in a Coulomb field \cite{bm}. The solution used Furry wave functions
in parabolic coordinates.  These radiative corrections reduced the
cross section by a constant amount \cite{bm,bm2},
\begin{equation}
\Delta\sigma(k) = - {28Z^2r_e^2\alpha\over9}f(Z)
\end{equation}
where $r_e = 2.8$ fm is the classical electron radius and
\begin{equation}
f(Z) = (\alpha Z)^2 \Sigma_{\nu=1}^\infty {1\over \nu(\nu^2+(\alpha Z)^2)}.
\label{fequation}
\end{equation}
For a heavy atom like lead, $f(82)=0.33$, and the Coulomb correction
reduces the cross section about 10\%, essentially independent of $k$.

For the differential cross section, the leading order correction
depends only slightly on $x$.  However the next term is large for
$x\rightarrow 0$ or $x\rightarrow 1$ \cite{leex}.  The sign of the
second term differs for the two extremes in $x$, introducing an
asymmetry between the electron and positron momentum spectra.  These
asymmetries have also been studied with perturbation theory; for heavy
nuclei, significant asymmetries have been predicted theoretically
\cite{brodsky} and observed experimentally \cite{ting}.

Tsai \cite{tsai} made detailed calculations of the screening factors
and radiative corrections for different materials.  He also considered
pair conversion on the atomic electrons in the target.  The
complexities of these calculations can be easily hidden by adjusting
$X_0$.  With this, for $Z > 4$ \cite{pdg},
\begin{equation}
X_0 = \bigg({4\alpha r_e^2 N_A\over A} \big[Z^2 [\ln{(184Z^{-1/3})} -f(Z)]
+Z\ln{(1194Z^{-2/3})}\big]\bigg)^{-1}.
\end{equation}
The first logarithmic term is for nuclear scattering, while the second
is for scattering from atomic electrons; $f(Z)$ is the Bethe-Maximon
correction.  For light nuclei with $Z\le 4$, the logarithmic terms are
inaccurate, and $Z-$dependent constants are usually used.  With this
$X_0$, Eqs. \ref{eq:constsigma} and \ref{eq:dsde} remain accurate and
easily usable.  At 10 GeV, these cross sections agree with the data to
within 2\% \cite{eickmeyer}.

At very high energies, other channels may contribute to the cross
section.  For example, the virtual photon in Fig. \ref{fig:scheme} can
be replaced with a virtual $Z^0$ boson.  Although the amplitude for
the $Z^0$ mediated reaction is not large, the interference with the
purely electromagnetic process can introduce $\approx 5\%$ asymmetries
between the electron and the positron, especially for large $M_p$
\cite{cahalan}.

There is a paucity of data on pair production from photons with $k>10$
GeV.  The high-energy frontier has moved to $e^+e^-$ and hadron
colliders, as will be discussed in Section 4.  There is been
considerable data on pair production in crystals, but the effects of
the crystal overshadow any subtleties in the individual photon-atom
interactions; crystals are discussed in Section \ref{s:fields}.

\section{Pair Production in Media}

As the photon energy rises, the formation length becomes large enough
that it encompasses more than one atom.  Then, the amplitudes for
interacting with multiple atoms must be added, and interference
between interactions with different atoms needs to be considered.

Landau and Pomeranchuk considered this problem in 1953
\cite{landau1,landau2}.  They used classical electrodynamics to study
bremsstrahlung.  The emission of photons with momentum $\vec{k}$ and
energy $k=|\vec{k}|$, from an electron starting with velocity $\vec{v}_1$ and
ending with velocity $\vec{v}_2$ into solid angle $\Omega$ is
\cite{jackson}
\begin{equation}
{d^2I\over dkd\Omega} = {Z^2e^2\over 4\pi^2} 
\ \ \bigg| 
{\vec{k}\times\vec{v_1} \over \vec{k}\cdot\vec{v}_1 - k} -
{\vec{k}\times\vec{v_2} \over \vec{k}\cdot\vec{v}_2 - k}\bigg|^2.
\label{eq:radiate}
\end{equation}

This equation is usually applied to an interaction of an electron with
a single atom.  Landau and Pomeranchuk realized that, when atoms are
close enough together, the interactions cannot be separated.  The
emission depends only on the initial and final velocities, and is
independent of any intermediate state velocities.  The scattering due
to nearby atoms adds together.

As long as the angle $\theta_e$ between $\vec{v}_1$ and $\vec{v}_2$ is
less than $1/\gamma$, where $\gamma=E/m$ is the Lorentz boost of the
electron, then the bremsstrahlung radiation is proportional to
$|\vec{v}_1-\vec{v}_2|^2$ and the radiation from the summed
independent (with uncorrelated directions) scattering is the same as
if the radiation from each scatterer was added independently.  In this
case, we define $\Delta\vec{v} = \vec{v}_1 -\vec{v}_2$, the electron
momentum change is $q=\gamma m |\Delta\vec{v}|$, and
\begin{equation}
{d^2I\over dkd\Omega} = 
{Z^2e^2\gamma^4 v^2\theta_e^2
\over \pi^2}
\ \ {1+\gamma^4\theta_e^4 \over (1 + \gamma^2\theta_e^2)^4}.
\label{eq:radiate2}
\end{equation}
Integrating over the solid angle gives the total emission:
\begin{equation}
{dI\over dk} = {2Z^2e^2q^2 \over 3\pi m^2}.
\label{eq:totalrad}
\end{equation}
Equation \ref{eq:totalrad} holds as long as the momentum transfer $q <
m$.  At larger momentum transfers, the radiation is reduced. The cross
section for classical bremsstrahlung can be found from
Eq. \ref{eq:radiate2}.

Landau and Pomeranchuk calculated the radiation from a long electron
trajectory that was determined by multiple scattering, with the
scattering spread evenly along the path.  They determined the
interference from radiation from different points on the path.
Radiation from nearby points added in-phase, while at larger
separations, the phase was random. The phase coherence holds for
shorter and shorter distances as the multiple scattering increased.
For energetic electrons radiating low energy photons, the reduction in
coherence reduces the total emission.

The conditions for reduced emission can be estimated from the multiple
scattering in a formation length.  When the multiple scattering angle
within $l_f$ is larger than $1/\gamma$, then the multiple scattering
reduces the coherence, and, with it, the radiation.  In a distance
$l_f$, the mean multiple scattering angle is \cite{pdg}
\begin{equation}
\theta_{MS} = {E_s \over E}\sqrt{l_f\over X_0}
\label{eq:ms}
\end{equation}
where $E_s = m\sqrt{4\pi/\alpha} = 21.2 $ MeV.  An additional
multiplicative term, $1+0.038\ln{(l_f/X_0)}$ is sometimes included to
account for the non-Gaussian nature of Coulomb scattering.  This
non-Gaussian nature is particularly apparent for thin targets, where
the number of scatters is small.  However, since Eq. \ref{eq:ms} is
only used for back-of-the-envelope calculations, this multiplicative
factor will not be used here.

Neglecting the non-Gaussian scattering tails, and requiring
$\theta_{MS} < 1/\gamma$, the emission of photons with energy $k$ from
electrons with energy $E$ is suppressed when
\begin{equation}
k < {E^2 E_S^2\over m^4 X_0}.
\end{equation}
In this limit, the bremsstrahlung spectrum is altered from the
Bethe-Heitler prediction $d\sigma/dk\approx 1/k$ to a harder spectrum,
$d\sigma/dk\approx 1/\sqrt{k}$!

Although pair production is not a classical process, Landau and
Pomeranchuk noted that similar reasoning should apply for it.  By
using crossing symmetry to relate bremsstrahlung and pair production,
they predicted that for photons with very high energies, the pair
production cross section scales as
\begin{equation}
\sigma \approx \sqrt{1\over k}
\end{equation}
This is very different from the Bethe-Heitler energy-independent cross
section.  At high enough energies, photons become penetrating
particles.

This cross section can be derived using the uncertainty principle
\cite{feinbergpomeranchuk} and relying on the proportionality between
the pair cross section and formation length.  The number of atoms in
the formation zone is proportional to $l_f$.  For coherent
interactions, the probability scales as the number of atoms squared.
The pair conversion amplitudes add in-phase over the formation length,
so the interaction probability scales as $l_f^2$, and the probability
per unit length scales as $l_f$.  Multiple scattering reduces $l_f$,
and with it the cross section.

Multiple scattering changes the electron and positron direction, but
not their momenta, reducing their longitudinal velocity.  The
longitudinal velocity reduction $\Delta v$ is
\begin{equation}
{\Delta v \over c} = \big[1-\cos(\theta_{MS})\big]\approx {\theta_{MS}^2\over 2}.
\end{equation}
The slowing reduces the longitunal momentum of the electron and
positron.  The excess momentum is transferred to the medium during the
scattering.  Including the multiple scattering from the positron and
electron, $\theta_{MS+}$ and $\theta_{MS-}$,
\begin{equation}
q_{||} = {m^2k\over 2E(k-E)} + {E\theta_{MS+}^2\over 2}
+{(k-E)\theta_{MS-}^2\over 2}
= {m^2k\over 2E(k-E)} + {E_s^2 l_f k \over 2E(k-E)X_0}
\label{eq:qp0}
\end{equation}
When $\theta_{MS}> 1/\gamma$ for the electron or positron, the
multiple scattering terms dominates.  This happens when 
\begin{equation}
l_{f0} > {m^2 \over E_s^2} X_0,
\label{eq:msmatters}
\end{equation}
{\it i.e.} when
\begin{equation}
k > {m^2M_p^2 X_0 \over 2\hbar E_s^2}.
\label{eq:kvsmp}
\end{equation}
In this regime, the multiple scattering angles depend on $l_f$, which
itself depends on $\theta_{MS}$.  This leads to a quadratic equation
for $l_f$.  When the multiple scattering dominates,
\begin{equation}
{\sigma\over \sigma_{BH}} = {l_f\over l_{f0}} =  \sqrt{kE_{LPM} \over E(k-E)} 
\approx {M_p\over m} \sqrt{E_{LPM}\over k}.
\label{eq:suppress}
\end{equation}
Here, $E_{LPM}$ is a material dependent constant, 
\begin{equation}
E_{LPM} = {m^4X_0 \over\hbar E_s^2} \approx 7.7\ {\rm TeV/cm}\cdot X_0.
\end{equation}
For lead, $E_{LPM} = 4.3$ TeV.  Sometimes $E_{LPM}$ is defined
differently, usually differing by a numerical factor (often 2).  Since
$E_{LPM}$ depends only on $X_0$, the pair production cross section may
be easily calculated for mixtures.

Pair production is suppressed for photons with $k> E_{LPM}$, {\it
i.e.} in the TeV region and above.  In the strong suppression limit
($k\gg E_{LPM}$), the formation length scales as $l_f\approx\sqrt{k}$,
and the cross section is suppressed as $l_f/l_{f0}\sim 1/\sqrt{k}$.
Suppression is largest for low-mass pairs.  This is not surprising;
the larger $M_p$, the shorter $l_{f0}$ (Eq. \ref{eq:lf0}), and the less
of an opportunity for multiple scattering to contribute to $q_{||}$.

While the pair production cross section falls at high energies, the
photonuclear cross section rises.  Photonuclear interactions occur
when the photon fluctuates into a quark-antiquark ($q\overline q$)
pair, which then interacts hadronically with a target nucleus.  Very
high energy photons may also interact directly with quarks in the
nucleus \cite{engel}.

\begin{figure}[bt]
\center{\includegraphics[angle=270,clip=,width=0.49\columnwidth]{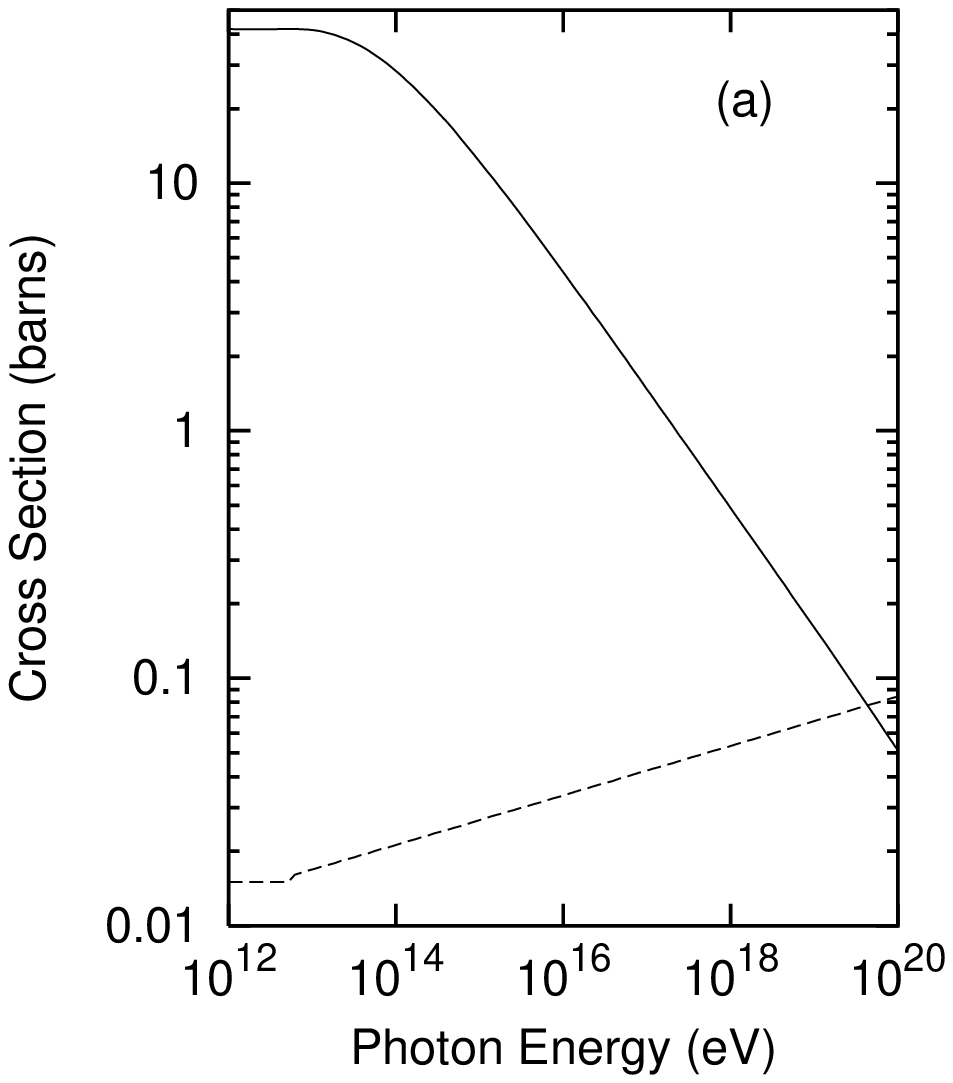}
\includegraphics[angle=270,clip=,width=0.49\columnwidth]{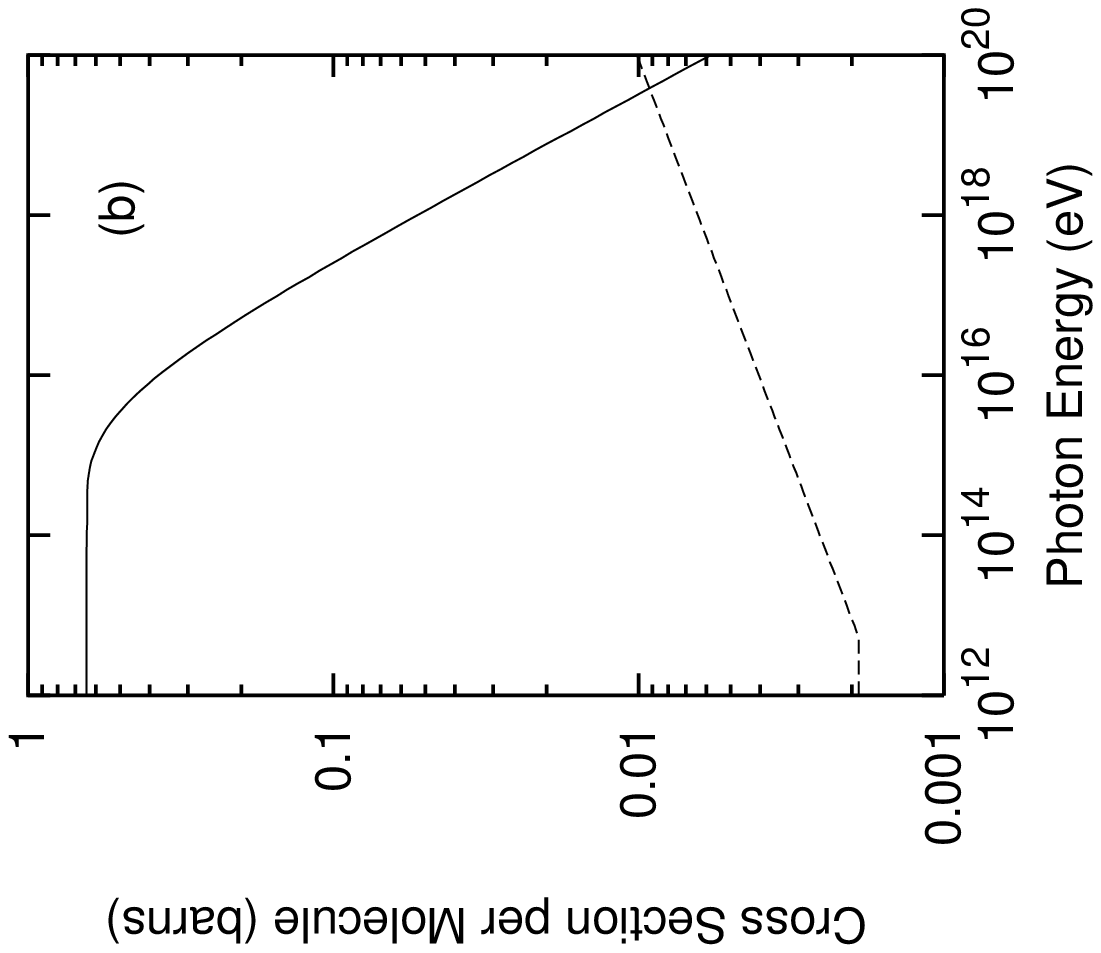}}
\caption[]{Cross sections for $\gamma\rightarrow e^+e^-$ (solid line)
compared with the cross section for photonuclear interactions (dotted
line) in (a) lead and (b) water.  Caveats for the curves are discussed
in the text.}
\label{fig:eehadron}
\end{figure}

When the photon energy is high enough, photonuclear interactions may
dominate over pair production \cite{ralston}.  Fig. \ref{fig:eehadron}
compares the pair production cross section in lead and water (using
Migdals suppression, see Sec. \ref{ss:migdal}) with the photonuclear
cross sections.  The photonuclear cross sections have not been
measured at these energies, but should increase slowly with energy.

For lead, the photonuclear cross section, $\sigma_{had}$ is about 15
mb when the photon-nucleon center of mass energy, $\sqrt{s_{\gamma
N}}$, is less than 100 GeV, and $15 (\sqrt{s_{\gamma N}/{\rm 100
GeV}})^{0.2}$ mb at higher energies.  For water, $\sigma_{had} = 1.9$
mb for $\sqrt{s_{\gamma N}} < 100$ GeV, and $1.9 (\sqrt{s_{\gamma
N}/{\rm 100 GeV}})^{0.2}$ mb at higher energies.  These
parameterizations roughly follow Fig. 5 of Ref. \cite{engel}; the
oxygen cross section is scaled from carbon as the nuclear cross
section, $A^{2/3}$, and the water cross section is the sum of its
constituent cross sections.  Any estimate in this energy range
requires a large extrapolation, so there is considerable uncertainty.
The energy-dependence in Ref. \cite{engel} is reasonably conservative.

There are also significant uncertainties in the pair production cross
section.  The cross section calculations may fail when the suppression
becomes very large (order of $\alpha\approx 1/137$).  When LPM
suppression is very large, higher-order reactions such as $\gamma
A\rightarrow e^+e^-\gamma A$ are more important.  With their larger
inherent $q_{||}$, these reactions are less affected by by multiple
scattering.

With these caveats, for both materials the photonuclear cross sections
are larger than the pair production cross sections for $k >
4\times10^{19}$ eV.  The crossover energy is similar for both
materials.  For lighter materials, $\sigma_{had}/\sigma_{BH}$ is
larger, while for heavier materials, LPM suppression is larger.  The
two effects largely cancel out.

Even with the uncertainties, the trend is clear: at very high
energies, photonuclear interactions become more important, and may
even dominate over pair production.  Even at a few $10^{17}$ eV,
electromagnetic showers may develop significant hadronic components,
complicating their identification.

Production of $\mu^+\mu^-$ and $\tau^+\tau^-$ pairs,
$\gamma\rightarrow l^+l^-$ is reduced by a factor $(m/m_l)^2$, where
$m_l$ is the lepton mass.  So, neither process is very important here.
Of course, in some parts of phase space (such as for high mass pairs),
the suppression is smaller.

One other suppression mechanism should be mentioned: bremsstrahlung
can also suppress pair production \cite{gg,skreview}. A produced
leptons can emit a bremsstrahlung photon and lose momentum.  When the
bremsstrahlung occurs in the pair production formation zone, the
combined reaction is a single, higher-order process.  the
bremsstrahlung will increase $q_{||}$, decreasing $l_f$, and with it,
the pair production cross section.

In a crude model, the kinematics changes when $l_f > X_0$.  This only
happens at very high energies, when the LPM effect is already
important. Including LPM suppression, $l_f > X_0$ when $k>E_s^2
X_0/\hbar$, or, for lead, $k > 4\times10^{19}$ eV.  This is only
approximate; while bremsstrahlung suppresses pair production, the
reverse is also true: pair production suppresses bremsstrahlung.  The
suppression of one reaction reduces its impact on the other reaction
\cite{ralston} and a combined calculation is needed to properly
understand pair production and bremsstrahlung in this energy regime.

The energies required to study LPM suppression of pair production are
well beyond the reach of existing accelerators.  However, LPM
suppression is important to studies of very high energy cosmic rays
(including neutrinos).  When high energy cosmic rays hit the
atmosphere, they interact, forming air showers which develop into
billions of lower energy particles.  Cosmic rays with energies up to
$3\times10^{20}$ eV have been observed \cite{nagano}.  If the incident
particles are photons (unlikely, as discussed in Section
\ref{s:fields}), then LPM suppression reduces the cross sections and
thereby lengthens the shower, increasing the amount of energy that
reaches the ground.  Proton induced showers are much less affected,
but still have an electromagnetic component that is subject to the LPM
effect.  However, the suppression is only relevant for the highest
energy showers \cite{skshowers}.  As will be discussed in Section
\ref{ss:experiment}, cosmic ray showers have been used to study the
LPM effect.

Very high energy cosmic ray $\nu_e$, $\nu_\mu$ and $\nu_\tau$ may be
visible at next generation neutrino detectors, like IceCube
\cite{icecube}. Neutrino interactions produce high energy leptons
along with hadronic showers from the struck nucleus.  $\nu_e$ showers
are most affected by the LPM effect, because of the very high energy
electron.  However, electrons from $\tau$ decay and the
electromagnetic component of the hadronic shower may also be affected
by LPM suppression.

\subsection{Thin Targets and Surface Production}

In very thin targets, the formation length may be larger than the
target thickness $T$. When this occurs, the target interacts as a
single unit.  Including LPM suppression, $l_f > T$ when
\begin{equation}
k > {2 mT\over \hbar} \approx 2.5\ {\rm TeV}\cdot T(\mu m).
\end{equation}
When $l_f > T$, LPM suppression may be less than in a very thick target.

For very thin targets, the multiple scattering in the target is
negligible, and the Bethe-Maximon cross section is retained,
independent of $k$.  The Bethe-Maximon cross section applies when the
$q_{||}$ from the multiple scattering (the second term in
Eq. \ref{eq:qp0}, with $T$ replacing $l_f$) is less than that due to
the pair production (the first term in Eq. \ref{eq:qp0}).  This occurs
for target thicknesses $T  <(m/E_S)^2X_0$.

For slightly thicker targets, $(m^2/E_S^2) X_0 < T <l_f$, the
conversion probability depends on the total scattering in the target.
This regime has been studied only for bremsstrahlung
\cite{shulgafomin}.  Assuming that the behavior is the same, for $k\gg
E_{LPM}$, the conversion probability scales as $\ln{(E_S^2T/m^2X_0)}$
- as the logarithm of the target thickness!

When $T > l_f $, the target no longer acts as a single unit. However,
there is an increased probability of interactions within $l_f$ of the
target surface.  The additional interactions are known as transition
pair creation, analogous to transition radiation.  The plasma
frequency of the medium, $\omega_p$ plays a key role. For solids,
$\hbar\omega_p$ is 30-80 eV, so about $\hbar\omega_p\approx 10^{-4}m$.
Taking $a=\hbar\omega_p/m$, \cite{bktransit}
\begin{equation}
P = {\alpha\over 2\pi} \big( {8a^2\over 35} +
{256a^3\over 256}\big),
\end{equation}
independent of the incident photon energy.  In a typical solid
$a\approx 10^{-4}$ and $P(a)\approx 10^{-10}$, so this effect is
generally negligible.

At higher energies, there are additional surface effects due to
multiple scattering.  A new type of transition radiation occurs as
the electrons electromagnetic fields rearrange themselves to account
for the multiple scattering in the medium.  Ternovskii calculated the
differential probabilities for bremsstrahlung and.  pair production in
plates of moderate thicknesses \cite{ternovskii}.  For plates that are
at least $T_0 = \alpha X_0/2\pi\xi$, ($1\le \xi\le 2$ rises slowly
with $k$ to account for the increasing shower length), the
differential probability for pair production is
\begin{equation}
{dN\over dk} = {\alpha\over2\pi k} 
\bigg(1+ \big[x^2 + (1-x)^2\big]\ln{({T\over T_0})}\bigg).
\end{equation}
The scaling with target thickness $T$ is far from linear!  This
equation only holds for relatively thin targets.  For thick targets,
surface transitions are treated as an add-on to production in the
bulk of the target.  Ternovskii gave formulae for surface transitions
which add-on to Migdals calculation of bulk transitions discussed in
the next section.

\subsection{Migdal Formulation for Pair Production}
\label{ss:migdal}

In 1956, Migdal published the first accurate calculation of
suppression of pair production \cite{migdal}.  He focused on
bremsstrahlung, but also provided formulae for pair production using
crossing symmetry.  He represented the transition probabilities with a
density matrix, and then averaged the density matrix over the possible
positions of the scattering centers.  This led to a Boltzman transport
equation for the radiation.  He solved this with the Fokker-Planck
technique.

This approach used the quantum mechanical matrix elements, with proper
account of electron spin, photon polarization and nuclear recoil.  The
scattering was realistically (randomly) distributed in space (albeit
with a Gaussian distribution for angles).

Migdal expressed his results in terms of two functions $G(\overline
s)$ and $\phi(\overline s)$, where $\overline s$ depends on the photon
and electron energies, and on the total radiation (Migdal used $s$ to
discuss bremsstrahlung, and $\overline s$ for pair production; the two
variables are closely related):
\begin{equation}
\overline s = \sqrt{({E_{LPM}k\over 8E(k-E)\xi(\overline s)})}
\label{eq:s}
\end{equation}
The factor $\xi(\overline s)$ increases from 1 to 2 as the pair
production moves from no suppression ($\overline s > 1$, the
Bethe-Heitler limit) to strong suppression ($\overline s\ll 1$):
\begin{eqnarray}
\xi(\overline s) & = & 1 \hskip 2.75 in (\overline s\ge 1) \nonumber \\
\xi(\overline s) & = & 1 + \ln(s)/\ln(s_1)          
\hskip 1.725 in (s_1 < \overline s < 1) \nonumber \\
\xi(\overline s) & = & 2 \hskip 2.75 in (\overline s < s_1) 
\end{eqnarray}
with $s_1 = Z^{2/3}/184^2$.  Migdal did not discuss mixtures of
materials.  However, the radiation is dominated by high$-Z$ materials,
it is not too far wrong to take the $Z$ of the heaviest atom in a
mixture. Alternately, one could, following the approach used for $X_0$
\cite{pdg}, take a weighted average, weighting the different materials
by their abundance times $Z^2$.

The equations for $\overline s$ and $\xi$
can be solved recursively.  However, $\xi(\overline s)$ varies very
slowly, and Stanev and collaborators found an approach that avoids the
recursion \cite{stanev}.  They defined $\overline s'$ following
Eq. \ref{eq:s}, with $\xi=1$.  Then,
\begin{eqnarray}
\xi(\overline s') & = & 1 \hskip 2.75 in (\overline s'\ge 1)
\nonumber \\
\xi(\overline s') & = & 1 + h - {0.08(1-h)[1-(1-h)^2]\over \sqrt{2}s_1}
\hskip 0.46 in (\sqrt{2}s_1 < \overline s' < 1) \nonumber \\
\xi(\overline s') & = & 2 \hskip 2.75 in (\overline s' < \sqrt{2}s_1) 
\end{eqnarray}
where $h=\ln{(\overline s')}/\ln{(\sqrt{2}s_1)}$.

The differential cross section for pair production is
\begin{equation}
{d\sigma\over dE} = {A \xi(\overline s) \over 3X_0 N_A k}
\bigg(G(\overline s) + 2\big[{E^2 + (k-E)^2\over k^2}\big]\phi(\overline s)
\bigg).
\label{eq:sigmamigdal}
\end{equation}

\begin{figure}[bt]
\center{\includegraphics[angle=270,clip=,width=0.8\columnwidth]{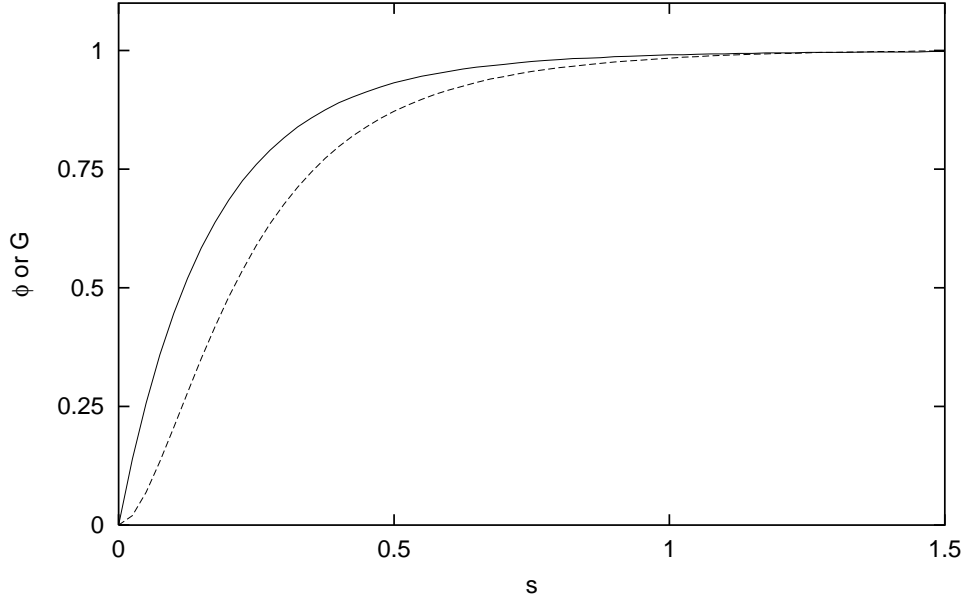}}
\caption[]{Migdals $G(\overline s)$ (dashed line) and $\phi(\overline
s)$ (solid line).}
\label{fig:fandg}
\end{figure}

Migdal gave infinite series for $\phi(\overline s)$ and $G(\overline
s)$.  However, they may also be represented as polynomials
\cite{stanev}:
\begin{equation}
\phi(\overline s) = 1 - \exp\bigg[-6\overline s
[1+(3-\pi)\overline s] + \overline s^3/(0.623+0.79\overline s
+0.658\overline s^2)\bigg]
\end{equation}
and
\begin{equation}
\psi(\overline s) = 1 - \exp[-4\overline s-8\overline s^2/
(1+3.96\overline s+4.97\overline s^2-0.05\overline s^3
+7.5\overline s^4)]
\end{equation}
with $G(\overline s) = 3\psi(\overline s)-2\phi(\overline s)$.
Figure~\ref{fig:fandg} shows these functions.

The limit $\overline s\rightarrow\infty$, $\xi(\overline s)=1$
corresponds to low photon energies, where $\phi(\overline s) =
G(\overline s) =1$ and the Bethe-Heitler cross section holds.  

The limit $\overline s\rightarrow 0$, $\xi(\overline s)= 2$
corresponds to strong suppression.  Then, $\phi(\overline s)=
6\overline s$, $G(\overline s) = 12\pi \overline s^2$ and the cross
section is
\begin{equation}
{d\sigma\over dE} = {4 A \over3 X_0 N_A k}
\bigg({E^2 + (k-E)^2\over k^2}\bigg)\phi(\overline s).
\end{equation}

\begin{figure}[bt]
\center{\includegraphics[angle=270,clip=,width=0.8\columnwidth]
{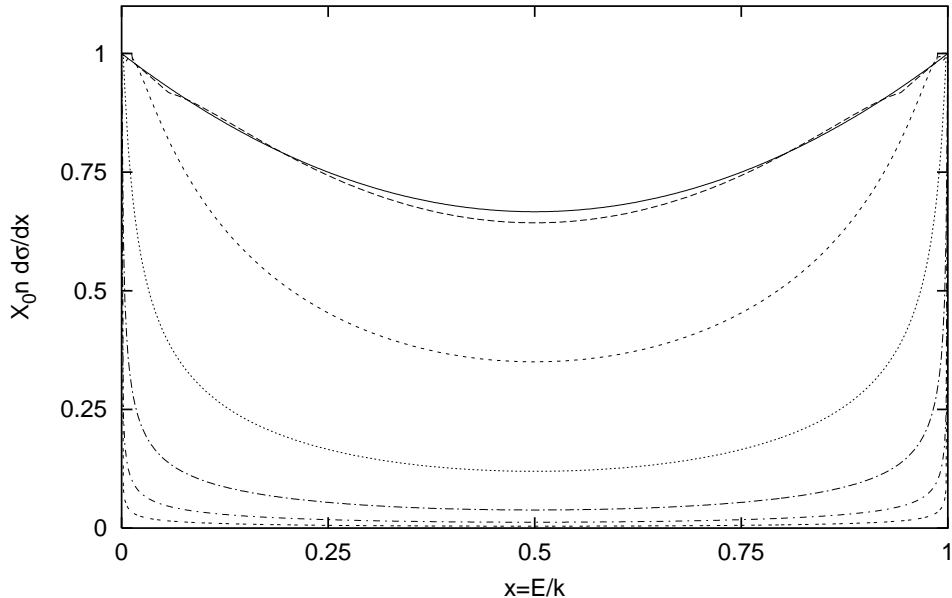}}
\caption[]{Differential cross section for pair production,
Eq. \ref{eq:sigmamigdal}, in a lead target, for different photon
energies.  Cross sections are shown for photons with $k=$ 1 TeV (top),
10 TeV, 100 TeV, 1 PeV, 10 PeV, 100 PeV and 1 EeV (bottom).  Here, $n$
is the number of atoms per unit volume.  When $E(k-E)\approx 1.2
kE_{LPM}$, the product $\phi(\overline s)\xi(\overline s)$ rises
slightly above 1 and the conversion probability for high energy
photons is slightly larger than at smaller $k$.  These curves can be
used for other materials by scaling $k$ by $E_{LPM}{\rm
(lead)}/E_{LPM}$; the bottom curve also applies for 3.2 EeV photons in
iron ($E_{LPM}= 13.6$ TeV).  }
\label{fig:migdalsigma}
\end{figure}

Figure \ref{fig:migdalsigma} shows the pair production cross sections
vs. $E/k$ in lead for different photon energies. At very high
energies, most conversions produce asymmetric pairs.

\begin{figure}[bt]
\center{\includegraphics[angle=270,clip=,width=0.8\columnwidth]{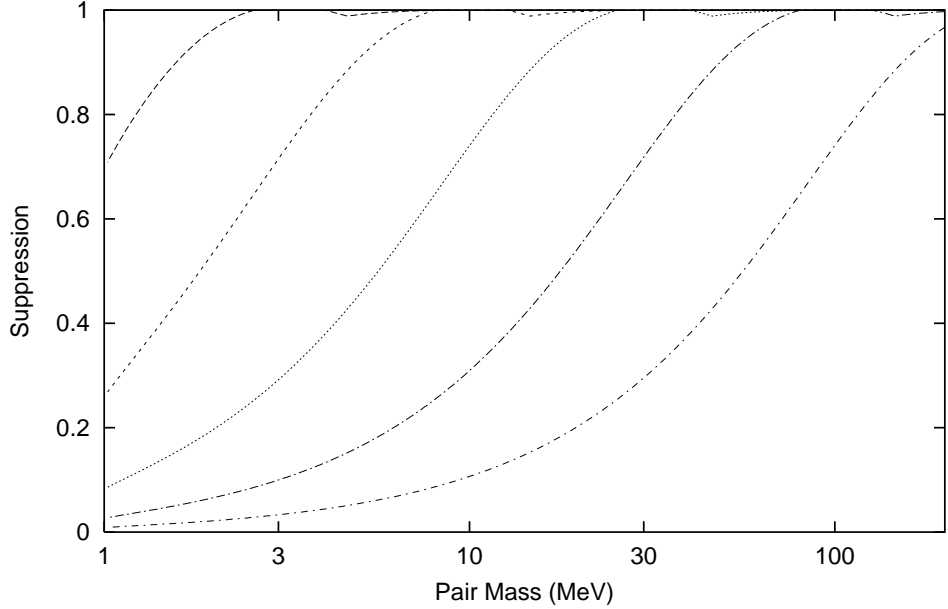}}
\caption[]{The suppression factor $S=\sigma_{LPM}/\sigma_{BH}$ as a
function of pair mass for photons with energies 44 TeV (top curve),
440 TeV, 4.4 PeV, 44 PeV and 440 PeV (bottom curve).  The small dips
below $S=1$ around $M_p^2\approx 8 m^2k/E_{LPM}$ are side-ripples to
the dips in Fig \ref{fig:migdalsigma}.  These curves can be applied to
other materials by scaling $k$ by $E_{LPM} {\rm (lead)}/E_{LPM}$; the
bottom curve also applies for 1.36 EeV photons in iron ($E_{LPM}=
13.6$ TeV).  }
\label{fig:compare}
\end{figure}

Figure \ref{fig:compare} gives the suppression (compared to
Bethe-Heitler) as a function of pair mass in lead for photons with
energies from 44 TeV to 440 PeV.  The suppression is largest for small
$M_p$. The Bethe-Heitler $d\sigma/dM_p$ shape is independent of $k$,
so the cross section is the product of this suppression factor with
the cross section from Fig. \ref{fig:bhwide}.  The bulk of the cross
section is near threshold, so the overall suppression is dominated by
the measurement around $2m\approx 1.12$ MeV.  This calculation
neglects the lepton transverse momenta, but most of the consequent
inaccuracy should cancel for the suppression.  Very roughly, in the
strong suppression limit, for $M_p\gg 2m$ the cross section scales as
$d\sigma/dM_p\approx 1/M_p^2$.

When the cross section suppression is large, the penetration depth of
electromagnetic showers increases significantly.  Not only is the
total cross section reduced, but either the electron or positron takes
most of the energy.

\subsection{Newer Calculations}
\label{ss:newer}

Several recent papers have presented more sophisticated calculations
of LPM suppression.  These calculations consider targets as an
integral whole, and therefore naturally handle finite target
thicknesses.  However, many authors considered only bremsstrahlung.
Often, too little information is given for an outsider to easily apply
the calculations to pair production.  They also do not consider
mixtures of atoms with different $Z$.  Blankenbecler and Drell used an
eikonal approach to study bremsstrahlung in finite targets
\cite{bland}.

Zakharov used a light cone path integral approach to study
bremsstrahlung.  He used a transverse Green's function to represent
$e^+e^-$ pair scattering from the atom and used it to solve the
Schroedingers equation.  He used an imaginary potential for the Greens
function, proportional to the cross section for an $e^+e^-$ pair to
scatter from the atom.

In the strong suppression limit, the scatterer is a dipole with length
$\rho$; the scattering cross section $\sigma(\rho)=C(\rho)\rho^2$
where $C(\rho)$ varies slowly with $\rho$ \cite{z1}.  With a constant
$C(\rho)$, this is Gaussian scattering. For an infinitely thick
target, the results are almost identical to Migdals.  With a slowly
varying $C(\rho)$ to model Coulomb scattering, the Zakharov cross
section has more energy dependence than Migdals $\xi(\overline s)$.
At higher energies, Zakharov should find slightly lower cross sections
than Migdal.

For regimes with moderate suppression, Zakharov adopted a more
accurate Coulomb potential than a dipole, with separate potentials for
the nucleus and it's electrons.  This separation is important for
low-$Z$ atoms where the electrons play a significant role \cite{z2}.
With these details, Zakharov found a good fit to the SLAC E-146 data,
except for the carbon.

V. N. Baier and V. M. Katkov made one of the few explicit calculations
of pair production \cite{bkone,bkpair}.  The framework was similar to
Migdal, but with a quasiclassical eikonal approximation for the
scattering.  This quasiclassical approach has been criticized as being
inapplicable in the regime where it was used because the electron
impact parameter cannot be treated classically \cite{z3}.  However,
Baier and Katkov argue that, when the angular momentum is large, their
approach is acceptable \cite{az3}.  The eikonal approach allowed them
to include Coulomb corrections in the potential, an important factor
in improving the accuracy of the calculations.  They included a
separate term in the potential to account for scattering from the
atomic electrons.  In the no-suppression limit, the authors obtain the
fully-screened Bethe-Maximon Coulomb-corrected cross sections.

Baier and Katkov note that, for heavy elements, the Coulomb correction
is larger than the uncertainties in the SLAC E-146 data
\cite{bkreview}.  The E-146 experimental analysis normalized the
Migdal calculations (without radiative corrections) to a radiation
length that included Coulomb corrections, effectively multiplying the
two effects together.  This was the limit of the then-current
technology.  Although the results matched the data, this may not have
been an optimal approach.

When the suppression is strong, Baier and Katkov give a polynomial
expression for the total pair production cross section
\begin{equation}
{\sigma\over \sigma_{BM}} \approx 4.28 \sqrt{E_{LPM}\over k}
\bigg[1-1.672\sqrt{E_{LPM}\over k} - 2.192{E_{LPM}\over k}
+{1\over 4L_1} \big(\ln{k\over4E_{LPM}} +0.274\big)\bigg]
\label{eq:bk}
\end{equation}
where $L_1= 183Z^{-1/3}e^{-f}$, with $f$ from Eq. \ref{fequation}.
Here, $\sigma_{BM}$ is the Bethe-Maximon cross section \cite{bkreview}
(apparently mis-labelled as Bethe-Heitler in Eq. 2.35 of
Ref. \cite{bkpair}).  The polynomial in Eq. \ref{eq:bk} reduces the
cross section at moderate photon energies.  However, at large enough
energies, the polynomial terms drop out, but the last correction term
in Eq. \ref{eq:bk} increases the cross section to 2-10\% above the
Migdal result.

\subsection{Experimental Measurements of Suppression}
\label{ss:experiment}

Suppression of pair production can only be observed for photon
energies $k>E_{LPM}$.  Since $E_{LPM} > $1 TeV, even for the densest
materials, it has not been possible to study this suppression at
accelerators.  Some studies have used cosmic rays, but the statistics
are limited by the high energy cosmic ray flux.  The earliest studies
considered low-energy electrons in photon induced showers.  Fowler,
Perkins and Pinau studied 47 electromagnetic showers with an energy
above 1 TeV in an emulsion stack, measuring the distance between the
initial conversion point and the first daughter pair downstream of the
primary \cite{fowler}.  At high photon energies, the average distance
rose, as expected for decreasing cross sections.  The statistics were
limited because few sufficiently energetic photons penetrate the
atmosphere.  To improve the statistics, some later experiments flew
emulsion detectors in high-altitude balloons \cite{balloons}, where
the flux of high-energy photons is higher.  They achieved somewhat
better statistics, but with qualitatively similar results.

One of the most precise studies was by Kasahara in 1985
\cite{kasahara}.  He studied the development of electromagnetic
showers with energies of order 100 TeV, in lead-emulsion chambers.  At
these energies, LPM suppression significantly increases the
penetration depth of the showers.  After removing the contamination
from hadronic showers, Kasahara found that the shower profiles were
consistent with Migdals cross sections, but not the Bethe-Heitler
formulae.

Because of the limited statistics possible in cosmic ray studies, the
best studies of the LPM effect have used accelerators to study the
suppression of bremsstrahlung.  Bremsstrahlung can be studied with
electron beams with energies $E\ll E_{LPM}$, as long as photons with
energies $k<E^2/E_{LPM}$ are studied.

The first accelerator experiment used 40 GeV electrons generated from
70 GeV protons at the Serpukhov U-70 accelerator \cite{varfolo}.
Bremsstrahlung photons from carbon, aluminum, lead and tungsten
targets were detected in a sodium-iodide calorimeter.  The electrons
were bent away from the calorimeter with a magnet.  The group studied
photons with energies between 20 and 80 MeV.  At higher photon
energies, the LPM effect was unimportant, while at lower energies, the
backgrounds were very high.  Even in the signal region, there were
significant backgrounds from bremsstrahlung in air and scintillation
counters, and muon contamination in the beam.  By taking ratios of
their data (lead/aluminum and tungsten/carbon), they were able to
estimate the degree of suppression.  They found suppression larger
than Migdal predicted, albeit with large errors.

A later experiment at the Stanford Linear Accelerator Center, SLAC
Experiment E-146, made precise measurements of bremsstrahlung from 7
targets, from carbon to uranium, in beams of 8 and 25 GeV electrons
\cite{e146prl1,e146prl2,e146prd}.  They studied photons with energies
between 200 keV and 500 MeV.  The photons were detected in a
position-sensitive segmented BGO calorimeter 50 meters away from the
target.  The small solid angle and position sensitivity helped reduce
backgrounds, especially from synchrotron radiation.  Data was taken at
2 different settings of calorimeter gain (i.e. photomultiplier tube
high voltage) to maximize the dynamic range.  The electron beam-line
was kept entirely in vacuum to eliminate bremsstrahlung from air or
target windows.

The group used a tertiary electron beam, but, because the primary beam
was also electrons, beam contamination was small. The beam was usually
run at an average of 1 electron per pulse, 120 pulses/second; events
containing zero or multiple electrons were rejected during analysis.

The calorimeter was calibrated using cosmic rays and a 500 MeV electron
beam. The group also used the downstream magnet for an electron
spectrometer, measuring the electron energy loss.  Because of the
limited resolution for electron energy loss, this was useful only as a
check.

\begin{figure}[p]
\center{\includegraphics[clip=,width=0.95\columnwidth]{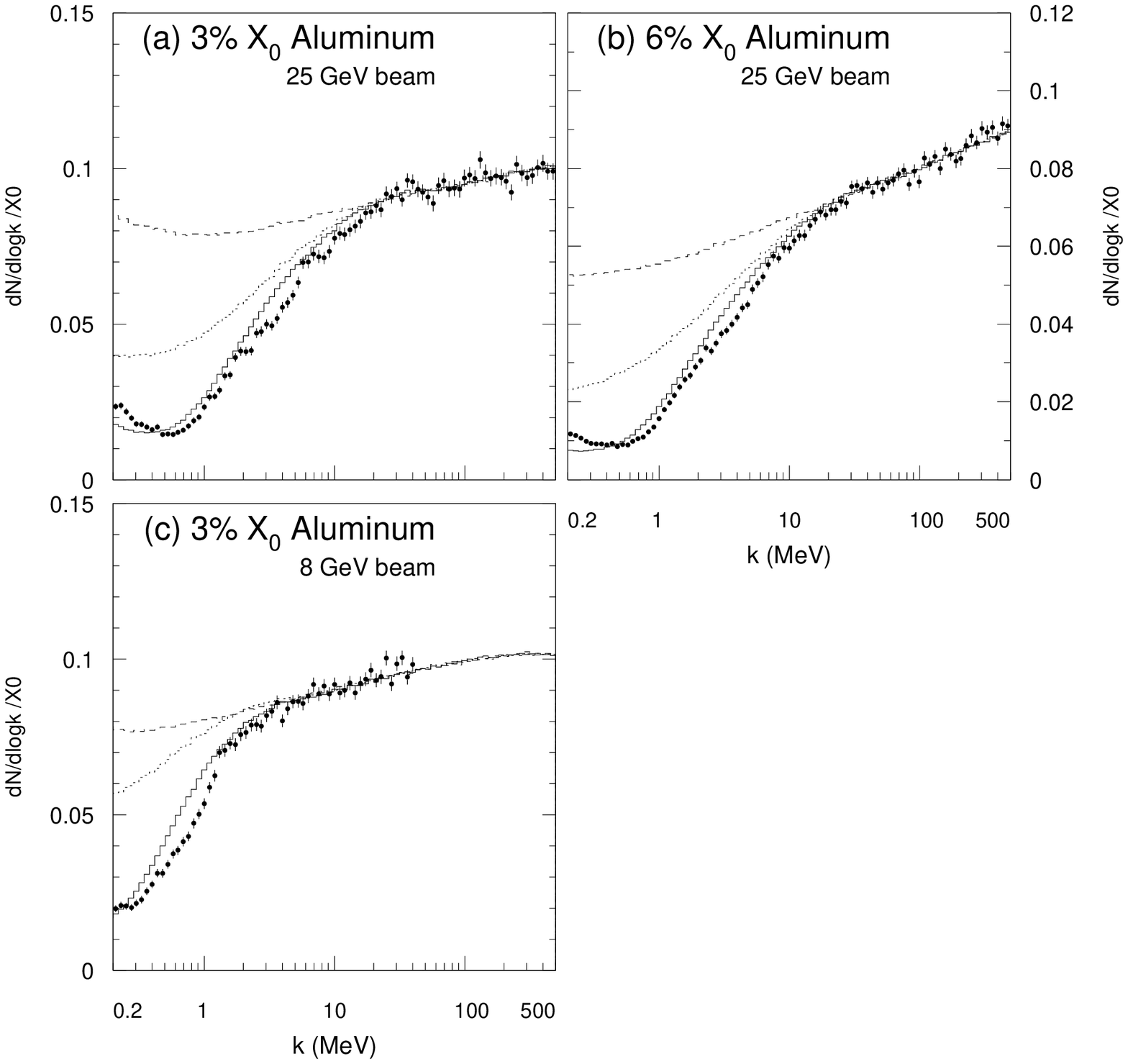}}
\caption[]{SLAC Experiment E-146 data (points) compared with Monte
Carlo predictions using the Bethe-Heitler (dashed line), LPM suppression
(dot-dashed line) and with LPM and dielectric suppression (solid line).
The simulations include transition radiation.  Adapted from Ref. 
\cite{e146prd}.}
\label{e146al}
\end{figure}

Figure \ref{e146al} shows the E-146 data for the aluminum targets.
Targets with thicknesses of 3\% and 6\% of $X_0$ (3.12 mm and 5.3 mm)
were used.  Events containing a single electron were selected, and the
photon energy spectrum measured.  The photon energy is plotted in
logarithmic width bins, with 25 bins per decade of energy.  Then, the
approximate Bethe-Heitler photon spectrum $dN/dk\approx 1/k$
transforms to $dN/d\ln{(k)}/X_0\approx 0.13$ - a straight line.  The
probability for a single electron to undergo two independent
interactions in the target tilts the spectra slightly.  This was dealt
with by using a Monte Carlo simulation.

The data shows significant suppression.  For $k < 2\times10^{-4}E$, an
additional suppression mechanism is required to explain the data.
This second mechanism involves the produced photon, which also
interacts with the medium, through forward Compton scattering.
Classically, this is described via a dielectric constant $\epsilon\ne
1$.

The group has also studied thin targets, where $T<l_f$ for relevant
photon energies.  For thin targets, the radiation was above the Migdal
calculation, and consistent with the calculations that treated the
entire target as a single radiator.

\begin{figure}[p]
\center{\includegraphics[clip=,width=0.6\columnwidth]{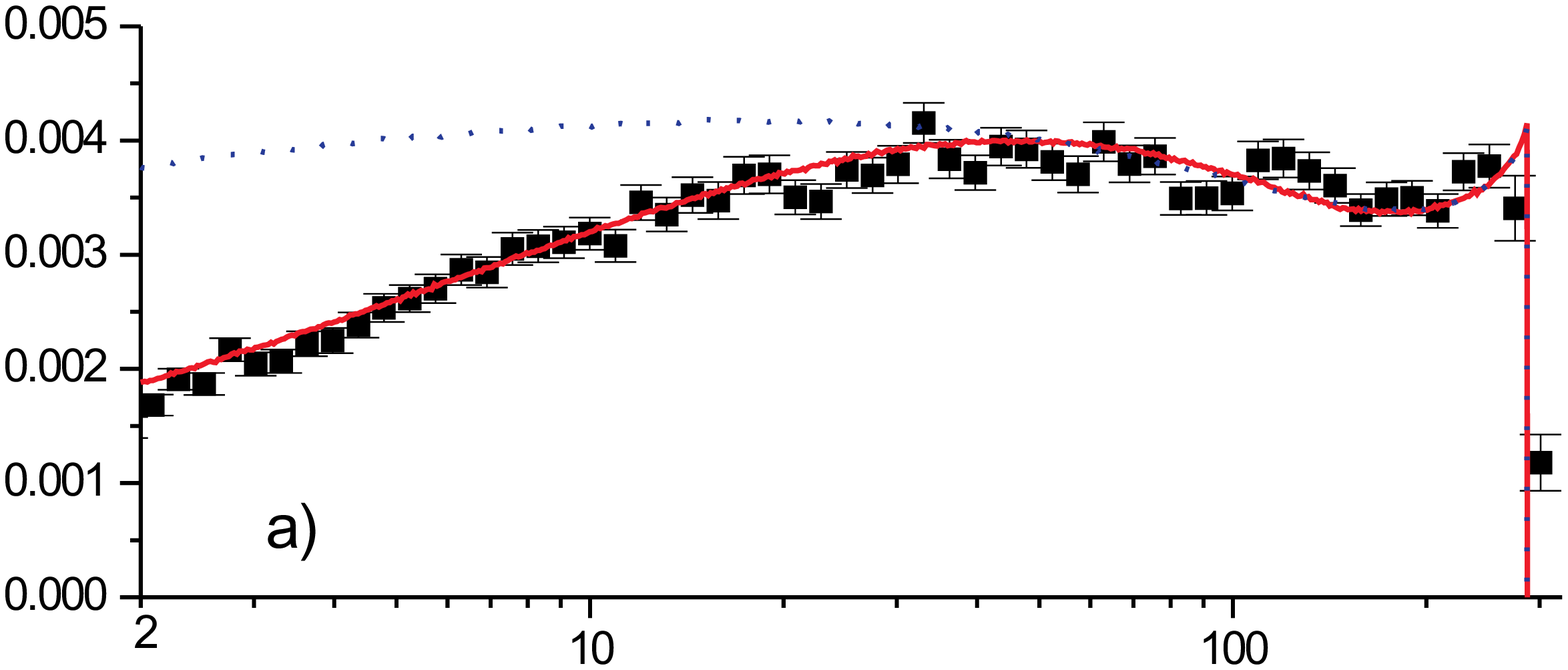}}
\center{\includegraphics[clip=,width=0.7\columnwidth]{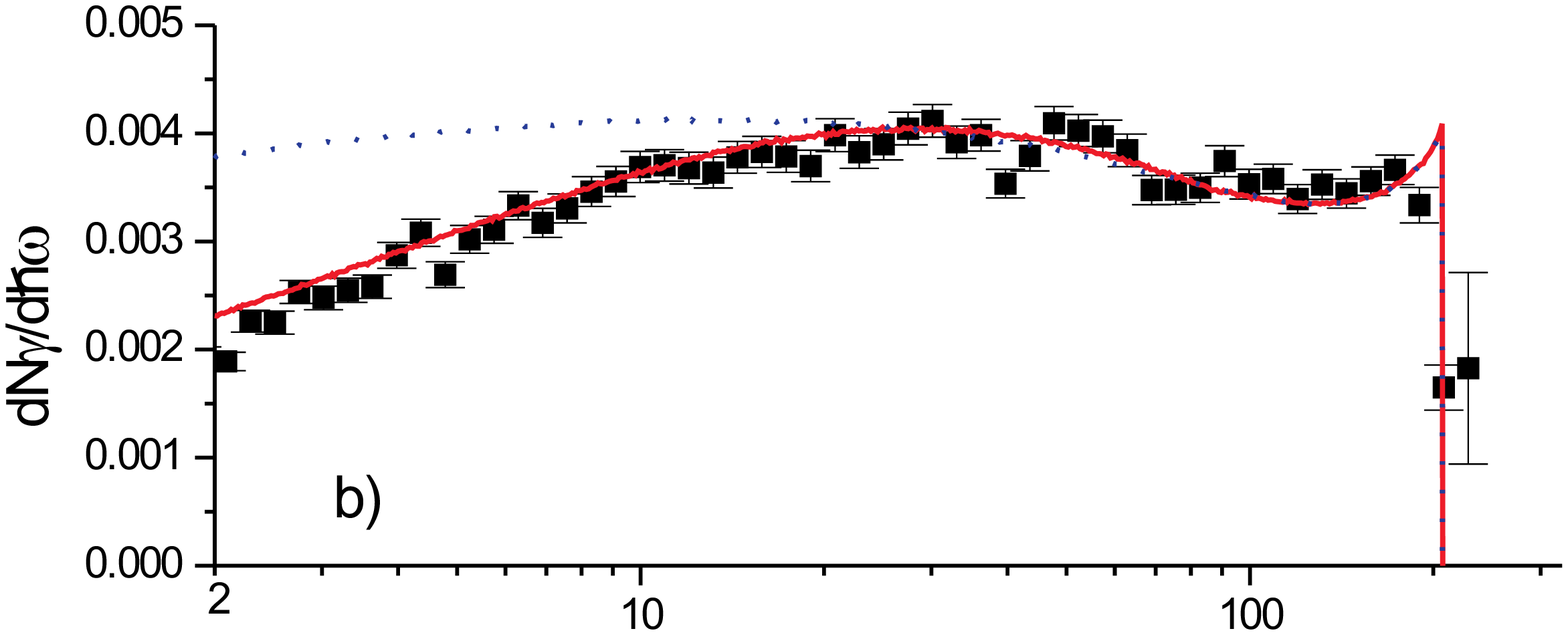}}
\center{\includegraphics[clip=,width=0.67\columnwidth]{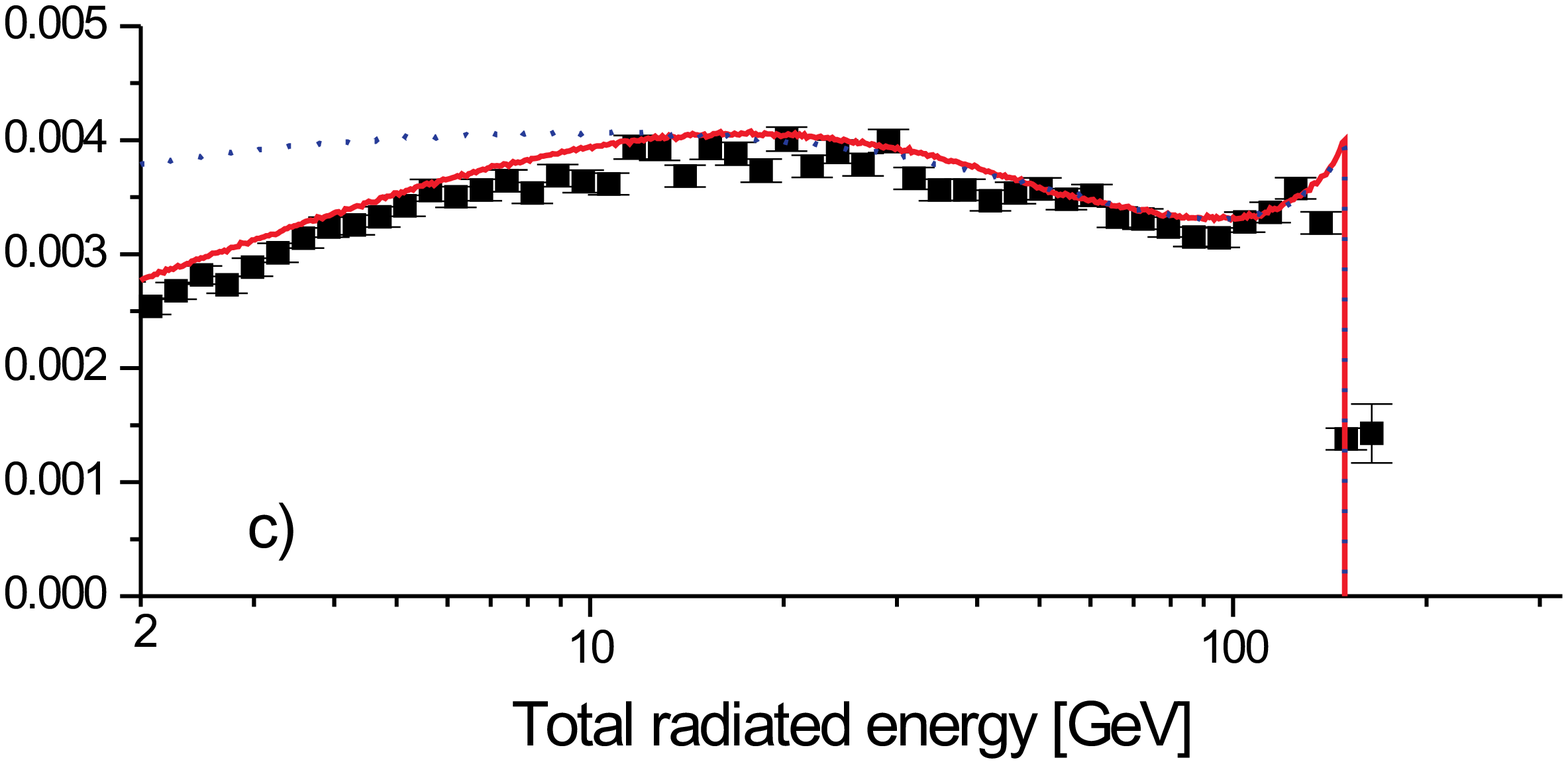}}
\caption[]{Bremsstrahlung spectrum from (a) 287 GeV, (b) 207 GeV and
(c) 149 GeV electrons striking a 0.128 mm thick iridium target.  The
photon energy scale is logarithmic, with 25 bins per decade if energy.
The dotted line is a simulation using the Bethe-Heitler spectrum,
while the solid line includes LPM suppression.  From
Ref. \cite{cern}.}
\label{fig:cern}
\end{figure}

More recently, an Aarhus-CERN-Florence-Free University, Amsterdam-Cape
Town collaboration studied bremsstrahlung from 149-287 GeV electrons
from iridium, tungsten and tantalum targets \cite{cern}.  They used a
CERN test beam.  Figure \ref{fig:cern} shows their tantalum target
data plotted in a manner similar to the E-146 data. With the higher
electron energies, they observed suppression up to $k/E\approx 0.16$,
a significant fraction of the spectrum.  The goal of this experiment
was to directly observe the overall increase in radiation length as
the radiation decreases.  They observed roughly a 30\% decrease in
radiation at the highest electron energy, consistent with expectations
from Migdals calculation.

The collaboration fitted their spectra to measure $E_{LPM}$ for their
targets \cite{cern2}. For the densest target, iridium, their measured
$E_{LPM}$ averaged $1.97\pm 0.16$ TeV, in quite good agreement with
the theoretical prediction of 2.25 TeV.  However, for the lighter
targets, tantalum and copper, the measured $E_{LPM}$ were considerably
below the expected values - by 40\% for copper.  The reasons for the
difference are not clear, but the collaboration suggests that possibly
bremsstrahlung from atomic electrons could be unsuppressed.  The
experiment also uses data from a carbon target for a normalization
spectrum; if the carbon spectra were somehow compromised, this data
could affect their measured $E_{LPM}$.  It may be worth noting that
the E-146 carbon-target data is poorly described by calculations based
on Migdals work.

\section{Pair Production in Constant Fields and Crystals}
\label{s:fields}

\begin{figure}[bt]
\center{\includegraphics[angle=270,clip=,width=0.8\columnwidth]{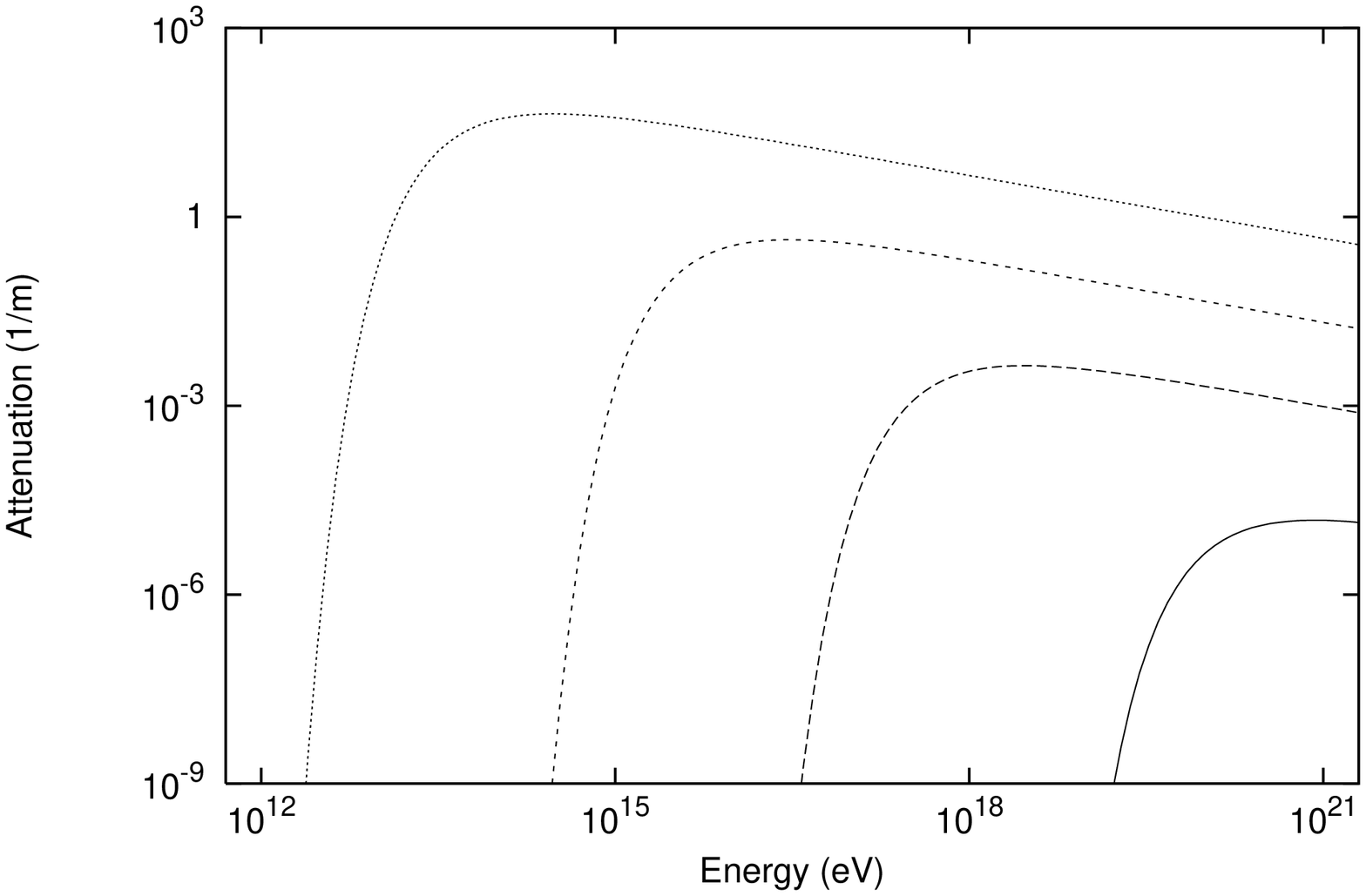}}
\caption[]{Attenuation coefficient (inverse attenuation length) for
photons in a magnetic field with perpendicular components of 0.35
Gauss (the Earth's average field - solid line), 100 Gauss (short
dashes), $10^4$ Gauss (long dashes) and $10^6$ Gauss (dotted line).}
\label{fig:bconvert}
\end{figure}

Over 50 years ago, H. Robl noted that for very high energy photons,
matter is not necessary for pair production: even static
electromagnetic fields suffice \cite{firstmag}.  This is analogous to
synchrotron radiation.  The photon conversion depends on the parameter
\begin{equation}
\chi = {kB_\perp\over 2mB_c}
\end{equation}
where $B_\perp$ is the magnetic field perpendicular to the photon
direction and $B_c = m^2/eh = 4.4\times10^{13}$ Gauss is the critical
field.  The probability of photon survival for a distance $d$ is
\begin{equation}
P(d) = \exp{(-a d)}
\end{equation}
where \cite{erber}
\begin{equation}
a = {0.16 \alpha m \over\lambda_e k} K^2_{1/3} (2/3\chi)
\label{eq:aconvert}
\end{equation}
and $K_{1/3}(x)$ is a modified Bessel function and $\lambda_e=386$ fm
is the Compton wavelength of the electron.  This equation is for the
constant field approximation, and may fail for rapidly varying fields.
Figure \ref{fig:bconvert} shows the attenuation coefficient $a$ for
photons of different energies.  The absorption is small for $\chi\ll
1$, or $k < 2mB_c/B_\perp$.  In this limit $a=0.46\exp{(-4/3\chi)}$.
The absorption rises sharply as a function of $k$ for $\chi > 1$,
reaching a peak for $k = 12 mB_c/B_\perp$.  At this maximum, $a\approx
0.1 (\alpha/\lambda_e) B_\perp/B_c$.  As $\chi$ continues to rise, the
attenuation coefficient drops slowly; for $\chi > 1000$, $a\approx
0.6\chi^{-1/3}\approx B_\perp^{2/3}k^{-1/3}$.  The attentuation
increases with increasing $B_\perp$, but, for fixed $B_\perp$, drops
slowly as $k$ rises.

There are some minor discrepancies between various studies.  Erber
\cite{erber} uses a dimensionless auxiliary function $T(\Upsilon)$ to
calculate $\alpha$.  His Eq. 3.3d, defines $T(\Upsilon)$, with
numerically matching small and large $\Upsilon$ limits in
Eq. 3.3c. However, Fig. 9 and Table VI of Ref. \cite{erber} give
$T(\Upsilon)$ are half as large as the equations; the equations appear
to be correct.  Ref. \cite{vankovpair} plots photon attenuation
following the equations in Erbers review.

The momentum transfer to the field required for pair production is the
same as for an atomic target, Eq \ref{eq:qperp0}.  Additional terms
are required to account for the magnetic bending which, like multiple
scattering, reduces the electron and positron longitudinal velocity.
The field bends electrons smoothly, with
\begin{equation}
\theta_B= {eB_\perp l_f \over E},
\end{equation}
in contrast to the $\theta\approx \sqrt{l_f}$ found for multiple
scattering.  Including the magnetic bending for the electron
($\theta_{B-}$) and positron ($\theta_{B+}$), the longitudinal
momentum transfer is
\begin{equation}
q_{||} = {M_p^2\over 2k} + {E\theta_{B-}^2\over 2} + {(k-E)\theta_{B+}^2\over 2}
={M_p^2\over 2k} + {e^2B^2_\perp l_f^2k \over 2E(k-E)}
\label{eq:qpb}
\end{equation}
This leads to a a quartic equation for $l_f$.  At low photon energies,
magnetic bending is small and $l_f$ rises linearly with $k$.  It
reaches a maximum, and then decreases .  For $\chi \gg 1$,
\begin{equation}
l_f = \bigg({\hbar  E(k-E) \over e^2B_\perp^2k}\bigg)^{1/3} 
=\bigg({\hbar k m^2 \over e^2B_\perp^2 M_p^2}\bigg)^{1/3}, 
\label{eq:qlf}
\end{equation}
so $l_f\approx k^{1/3}/B_\perp^{2/3}$ - the same scaling found from
the more detailed calculations.

Since $B_c$ is so large, this process is only important for photons
with quite high energies.  In fact, conversion in an external magnetic
field has not been observed.  However, it does have important
astrophysical applications.  One example involves very high energy
cosmic rays.  Very high energy photons may pair convert in the earth's
magnetic field \cite{vankovpair,mcbreenpair,stanevpair}.  For a $k =
10^{20}$ eV photon perpendicular to the earth's field, the attenuation
length is about 100 km.  The earth's magnetic field extends to
altitudes of 1000's of kilometers, far above the atmosphere where
lower energy showers originate.  The produced electrons interact with
the earth's magnetic field and emit synchrotron radiation.  The
combination of pair production in the field and synchrotron radiation
leads to an electromagnetic shower, even in the absence of matter.
The field-induced shower largely stops developing when the average
particle energy drops so that $\chi \ll 1$ ($\approx 10^{19.5}$
eV). The produced particles propagate downward until they reach the
atmosphere, where the shower develops conventionally.  The initial
high-altitude reactions gives these showers a head-start.  This early
development is helpful in differentiating photon initiate showers from
proton initiated showers.  Only a handful of showers with energies
above $10^{20}$ eV have been observed, but the primary particles
cannot all be photons \cite{nagano}.

Pair conversion can also occur in the strong magnetic fields that
surround many astrophysical objects.  For example, many neutron stars
support surface magnetic fields around $10^{12}$ G.  In these fields
pair conversion occurs quickly for $k>50$ MeV.

Although pair conversion from an external magnetic field has not been
seen in the laboratory, a fairly close analog has been studied: pair
conversion in crystals.  When a photon traverses a crystal parallel to
one of the axes, the magnetic fields of the individual atoms add
together to act as a single, extremely strong field.  The effective
fields are largest when the photon direction is well aligned with a
crystal axis, within a characteristic angle given by the atomic
lattice spacing divided by the formation length.  For 100 GeV photons,
this angle is a few mrad.  Often, the target is mounted in a
goniometer which allowed precise adjustments of the crystal
orientation.  Pair conversion is studied as the crystal orientation is
varied; when the axis is properly aligned, pair production is enhanced
\cite{moore}.

Two factors affect the enhancement.  Impurities and other lattice
defects lead to uneven atom spacings which reduce the coherence; high
quality crystals are required.  The quality requirement limits the
choice of target material.  At the highest photon energies, variations
in atomic spacing due to lattice vibrations limit the coherence.
Sometimes, crystals are cooled to reduce these vibrations. The second
factor is the divergence (angular size) of the photon beam.  In a beam
with a large divergence, many photons will not be within the
characteristic angle of the crystal axis.

The simplest manifestation of the strong coherent fields is
channelling; charged particles in crystals may be strongly deflected.
Channeling studies show that the effective fields can reach 25
Megagauss in silicon crystals \cite{channel}.  Crystal channeling can
in some cases supplant conventional accelerator magnets, particularly
for beam extraction \cite{guidi}.

With the increased bending, production of bremsstrahlung photons is
also greatly enhanced. This technique is mature enough that plans are
advanced to use coherent bremsstrahlung in a diamond crystal to
generate high-energy polarized photon beams for a series of
experiments \cite{bosted}.

The fields can also cause photon conversions.  At CERN, pair
conversion in germanium, tungsten and iridium crystals has been
studied with photons with energies up to 150 GeV.  The pair production
cross section was enhanced over the isolated atom Bethe-Heitler
prediction; the enhancement grew linearly with energy, reaching a
factor of 7 for $k=150$ GeV \cite{kirsebom}.  The total and
differential (with respect to $E/k$) cross sections agreed with the
predictions for coherent enhancement.

When a magnetic field also contains matter, both fields must be
considered together.  The effects of the nuclei and the fields must be
combined to find the conversion probability.  This is a complex
calculation.  Here we examine one limit, the reduction in pair
conversion on a target due to the motion caused by a moderate ($\chi <
1$) magnetic field. Here, the magnetic field plays a similar role to
multiple scattering.  When the magnetic bending terms in
Eq. \ref{eq:qpb} dominate, the magnetic field suppresses the
bremsstrahlung.  The suppression equation is quartic, and, in the
limit of strong suppression, the cross section is \cite{skreview}
\begin{equation}
\sigma=\sigma_{BH} \bigg({kE_B\over E(k-E)}\bigg)^{2/3}
\end{equation}
where 
\begin{equation}
E_B={m B_c\over B_\perp }
\end{equation}
plays a similar role to $E_{LPM}$.  This approach only applies for
very high energy photons in relatively low-density matter.  At high
densities, LPM suppression dominates.  When $\chi >1$, pair conversion
in the magnetic field is important. A parallel analysis finds a
similar suppression for bremsstrahlung. The bremsstrahlung calculation
is supported by detailed calculation using kinetic equations
\cite{bks}.

\section{Pair production at particle colliders}

Pair production also occurs at particle colliders.  The colliding
particles each emit a photon; the two photons collide to form an
$e^+e^-$ or other lepton pair.  Lepton pair production at $e^+e^-$
colliders has been studied intensely over the past 30 years.  These
studies have been used for a variety of tests of QED \cite{budnev}.
The reactions have also been studied as backgrounds to other
interesting processes, especially those that are beyond the standard
model.  They are also used to measure the internal (electromagnetic)
structure of the photon \cite{leptons}.

Although these processes follow the same diagrams as pair production,
Fig. 1, there are some significant differences from the fixed target
regime.  The reaction $e^+e^-\rightarrow e^+e^-e^+e^-$, studied at
$e^+e^-$ colliders, involves many more diagrams than appear in Fig. 1.
Second, the photons are virtual.  For $e^+e^-$ colliders, and $ep$
colliders, the virtuality (effective mass) is large, while in proton
and ion colliders it is smaller.  This review will focus on
photoproduction with almost-real photons (those with small virtuality)
at $ep$, $pp$/$\overline pp$ and heavy ion colliders.  Finally, the
symmetric initial state leads to some interesting effects. There is a
vast literature on photoproduction at colliders; here we only consider
a few selected examples to illustrate the similarities and differences
with fixed target reactions.

Muon and electron pair production, $ep\rightarrow epe^+e^-$ and
$ep\rightarrow ep\mu^+\mu^-$ were studied by the H1 collaboration at
the HERA $ep$ collider \cite{leissner}.  They observed lepton pairs
with masses up to 80 GeV, where electroweak corrections to QED are
significant.  The study was done to search for physics beyond the
standard model, but the collaboration found good agreement with the
standard model.

In colliders, the ions are fully stripped of electrons.  There is no
screening, and, as the beam energy rises, photoproduction occurs at
larger and larger impact parameters.  Consequently, the cross section
continues to rise with energy indefinitely.  For example, in gold on
gold collisions at a center of mass energy of 200 GeV per nucleon at
Brookhaven's Relativistic Heavy Ion Collider, the cross section is
expected to be 33,000 barns, rising to 200,000 barns in 5.5 TeV per
nucleon lead on lead collisions at CERN's Large Hadron Collider
\cite{baurreview}.  Pair production can occur at ion-ion separations
up to several $\mu$m.

Because $Z\alpha \approx 0.6$, higher order corrections to the cross
section could be large, and pair production may probe non-perturbative
strong-field QED.  This possibility encouraged many theoretical
efforts.  One approach extended the Bethe-Maximon approach to
collisions of heavy ions. The effect of the Coulomb potentials of both
ions was to reduce the cross section by 25\% at RHIC and 14\% at the
LHC \cite{ivanov}.

Tony Baltz made an all-orders calculation of pair production in
ultra-relativistic heavy ion collisions.  He found an exact solution
to the time-dependent Dirac equation \cite{baltz}. In an appropriate
gauge, the Coulomb potential from an ultra-relativistic particle
simplifies into a simple two-dimensional potential, and pair
production can be calculated exactly.  Surprisingly, his results
matched the lowest order perturbative result \cite{baltz}, without any
Coulomb corrections.  However, a later, more careful calculation found
results that included the Coulomb corrections \cite{lee}.  The
theoretical issues are discussed elsewhere in this volume, so they
will not be considered further here.

In most existing data on $e^+e^-$ production in ion collisions, at
least one of the ions is light (typically sulfur).  Experiment NA-45
at the CERN SPS studied 200 GeV per nucleon bare sulfur ions hitting a
platinum target and found cross sections consistent with the lowest
order perturbative result \cite{prev2}.  The measured cross section in
a restricted kinematic region is $0.99^{+0.33}_{-0.28}$ of that
expected from lowest order QED.  Here, the errors are combined in
quadrature.  This result could be used to put limits on some models of
radiative corrections.  Other SPS studies found similar results, with
general agreement to lowest order calculations \cite{prev1,prev3}.

At ion colliders, the charges are large enough that multiple
interactions between a single ion pair are very possible.  For
example, two pairs may be created by a single ion-ion reaction
\cite{pairpair}.  Multiple pairs have not been studied yet.  However,
the STAR collaboration has studied $e^+e^-$ pair production
accompanied by mutual Coulomb dissociation of the two nuclei
\cite{vladimir,mylowx}.  The mutual Coulomb dissociation criteria
selects collisions where the impact parameter is less than about 30
fm, considerably smaller than for unselected pair production.  The
reduction in impact parameter leads to stronger colliding fields, and
therefore to increased higher order corrections.  Despite this, the
collaboration found that their results were well described by lowest
order QED.  They did find that it was necessary to include the photon
virtuality in the calculation to explain the $p_T$ spectrum of the
pair.

Multiple interactions are most common at small impact parameters,
where the field densities are strongest.  For this reason, they are a
good place to search for higher order effects.  Despite this, the STAR
data are consistent with lowest order QED.  Because of the restricted
impact parameter range, this data cannot be directly compared with 
existing studies of radiative corrections \cite{mylowx}.

The absence of atomic electrons allows some new reactions, such as
pair production with capture, where an electron is produced bound to
one of the participating nuclei.  This reaction has a number of
notable applications, most notably the first production of
antihydrogen.  Antihydrogen is produced when positrons were
produced via pair production, bound to antiprotons
\cite{antihydrogen}.  At heavy ion colliders, the cross section for
electron capture is sizable, of the order of 100-200 barns at RHIC and
the LHC \cite{baur2}.  The single-electron atoms are lost from the
colliding beams, reducing the luminosity, and giving pair production
some importance for accelerator design.

\section{Conclusions}

Pair production at high energies covers a varied palette of physics.
High mass pairs have been a sensitive test of QED, and pair production
at $e^+e^-$ and $ep$ colliders continues to be a laboratory to search
for new physics.  These colliders allow the highest energy tests of
QED.

Lower mass pairs have long formation lengths.  These reactions are
distributed over many atoms, and probe bulk characteristics of the
material.  When the photon energy is large enough, pair production is
suppressed below the Bethe-Heitler cross section, and many `obvious'
scaling laws fail.  At very high photon energies, the pair production
cross section may be smaller than the photonuclear cross section, and
photonic showers will look like hadronic showers.

Pairs may also be produced in strong magnetic fields.  This has
significant implications in astrophysics.  These strong magnetic
fields also appear in crystals when the photon is aligned with one of
the crystal axes.  These aligned photons may coherently convert to
$e^+e^-$ pairs.

I would like to thank Ulrik Uggerh\o j for permission to reprint
Fig. \ref{fig:cern} and for useful discussions.  Kai Hencken, Gerhard
Baur, Roman Lee and Hristofor Vankov all provided useful suggestions.
This work was supported by the U.Sy. Department of Energy under
Contract No. DE-AC-03076SF00098.

\end{document}